# The properties of plasma sheath containing the primary electrons with a Cairns-distribution

Yida Zhang, Jiulin Du

*Department of Physics, School of Science, Tianjin University, Tianjin, China*

**Abstract** We study the properties of plasma sheath containing the cold positive ions, the secondary electrons, and the primary electrons with a Cairns-distribution (a non-thermal velocity-distribution). We derive the generalized Bohm criterion and Bohm speed, the new floating potential at the wall, and the new critical secondary electron emission coefficient. We show that these properties of the plasma sheath depend significantly on the α-parameter in the non-thermal α-distribution, and so they are generally different from those of the plasma sheath if the primary electrons were assumed to be a Maxwellian distribution.

---

## 1. Introduction

When plasma is in contact with other objects, a narrow space positive charge region is formed near the contact interface. This region is called the plasma sheath, and thickness of it is about ten Debye lengths. In the field of materials industry, study of the plasma sheath plays an extremely important role for guiding improvement of material surface modification and etching technology. Meanwhile, the relevant theories of plasma sheath have been widely applied to magnetic fusion devices, Hall thrusters and other equipment [1-3].

Analyzing the Bohm criterion is a common way to study the properties of plasma sheath. If ions could enter the sheath region, they need to satisfy the condition of minimum ion drift velocity, and a criterion based on this condition is called the Bohm criterion. In other words, the drift velocity of ions that can enter the sheath region must be greater than a certain minimum speed, which is called Bohm speed. The Bohm criterion can not only be used to judge stability of the plasma sheath, but also to study other properties of the sheath. Early in 1974 [4], it was pointed that when the plasma sheath is formed in an electron-ion plasma, the drift speed of ions entering the sheath was no less than the ion sound speed.

Due to the complexity of plasma application environments, plasma can have different compositions in different situations, such as the presence of negative ions, secondary electrons and so on. Usually, the case is present in the plasma sheath that there is a charging phenomenon by the collisions between electrons and other objects such as walls, workpieces, and dust particles etc. Electron bombardment to other objects will lead to secondary electron emission (SEE), and thus the sheath contains the secondary electrons. The plasma compositions can affect properties of the sheath, such as the Bohm criterion etc. It was shown that [5] electrons produced by SEE can significantly change the potential distribution of the sheath. The intensity of SEE is described by the SEE coefficient $\gamma$, defined as

$$\gamma = \frac{\sigma_s}{\sigma_p} . \tag{1}$$

where $\sigma_s$ is SEE flux, and $\sigma_p$ is primary electron flux.

In addition to the fact that plasma components can change the sheath properties, the velocity distribution of charged particles also change the sheath structure. Traditionally, most of the studies were based on the assumption that electrons in the plasma are a Maxwellian velocity distribution. However, due to the possible existence of non-thermal electrons, the velocity distributions often deviate from the Maxwellian one, which requires us to analyze the new properties of the plasma sheath with non-Maxwellian velocity distribution. In recent years, various non-Maxwellian velocity distributions have been observed and applied to study new properties in nonequilibrium complex plasmas, such as the kappa-distribution, the nonextensive q-distribution and the Cairns distribution etc. Sharifian et al.[7] discussed the non- extensibility of electrons to study the floating potential of the plasma sheath. Hatami et al. used two types of sheath models with the nonextensive q-distributed electrons and thermal ions to study the sheath structure and Bohm criterion[8, 9], and later studied the Debye length in the magnetic sheath model [10]. Borgohain et al. studied the plasma sheath with the nonextensive q-distributed two temperature electrons [11] and the electronegative plasma sheath with the nonextensive q-distributed electrons [12]. Zhao [13] et al. studied the effect of the plasma sheath with the nonextensive q-distributed electrons on the SEE. In addition, Dhawan et al.[14] considered the q distribution of electron, ion temperature and ion neutral collision in the simulation. Ghani [15] et al. studied the dusty plasma sheath structure with superextensive electron and SEE effect. Bouzit et al. studied the sheath formation with the electrons following a Cairns-Tsallis distribution [16], and Asserghine et al studied effect of SEE on the plasma sheath with the nonextensive q-distributed electrons [17], etc.

Among the many models with non-Maxwellian distributions in complex plasmas, the non-thermal $\alpha$-distribution (i.e., the Cairns distribution) is a typical non-equilibrium model that has been widely used to study new properties of the plasmas. As known, the non-Maxwellian velocity distribution of primary electrons significantly affects the secondary electron emission (SEE) flux[18] and consequently alters the structure and properties of the plasma sheath. In this regard, the Cairns distribution offers a physically distinct and complementary description for nonthermal electrons. Unlike Tsallis-type distributions, which mainly modify the asymptotic high-energy behavior, the Cairns distribution introduces a controlled nonthermal population at finite velocities through the parameter $\alpha$, resulting in a non-monotonic velocity profile. This characteristic has been shown to modify sheath properties considerably even when the high-energy tail remains close to a Maxwellian one. For example, the fluid dynamics as well as numerical analyses of electrostatic plasma sheaths containing nonthermal electrons revealed that adopting a Cairns-type distribution leads to substantial changes in sheath potential, charge density profiles, and the Bohm criterion, particularly in the presence of multi-component ions, dust particles, or external fields[24]. Similar findings were also reported in the numerical studies of electrostatic sheaths with charged nanoparticles[25], where nonthermal electrons were found to enhance intermediate-energy transport and strongly influence sheath structure and stability.

Motivated by these results and the growing evidence that edge and near-wall plasmas may exhibit nonthermal features beyond simple power-law tails, in this work we re-examine the fundamental sheath properties, including the Bohm velocity, the floating potential and the secondary electron emission by using the Cairns distribution. Our approach is to provide an alternative and complementary framework suitable for regimes where intermediate-energy nonthermal populations play a dominant role. In this study, we examine several properties of the plasma sheath in the presence of non-thermal primary electrons with the Cairns distribution.

The paper is organized as follows. In Sec.II, we introduce the basic theory of plasma sheath and the Cairns $\alpha$-distribution. In Sec.III, we study properties of the plasma sheath if the primary electrons are the Cairns $\alpha$-distribution. In Sec.IV, we make numerical analyses, and in the Sec.V, we give the conclusions.

## II. The basic theory of plasma sheath and the Cairns-distribution

For the convenience of discussion, we consider a plasma sheath to be in a stable state, the ions to be cold, and the plasma model to be a one-dimensional non-collision form. The boundary between the sheath and the neutral region of the plasma is as the starting point of one-dimensional coordinate, $x$=0, and the position of the wall is set as $x=x_w$ (See Figure 1).

If $v_s$ and $v_i$ are the secondary electron velocity and the ion velocity respectively, $n_s$ and $n_i$ are the number density of the secondary electrons and ions respectively, $e$ is the elementary charge, and $\varphi$ is the electrostatic potential function, one can have the following hydrodynamics model of the plasma sheath [9],

$$m_e v_s \frac{\partial v_s}{\partial x} = e\frac{\partial \varphi}{\partial x}, \tag{2}$$

$$\frac{\partial (n_s v_s)}{\partial x} = 0, \tag{3}$$

$$m_i v_i \frac{\partial v_i}{\partial x} = -e\frac{\partial \varphi}{\partial x}, \tag{4}$$

$$\frac{\partial (n_i v_i)}{\partial x} = 0. \tag{5}$$

By combining Eq. (4) and Eq. (5), one obtains the number density of ions,

$$n_i = n_{i0}\sqrt{\frac{m_i v_{i0}^2}{m_i v_{i0}^2 - 2e\varphi}}, \tag{6}$$

where $m_e$ and $m_i$ are the mass of an electron and an ion, $n_{i0}$ and $v_{i0}$ are respectively the number density and velocity of ions at the sheath boundary $x$=0.

If one uses the subscript $j$=$s$, $p$, and $i$ to represent the secondary electrons, the primary electrons and the ions, respectively, $\sigma_{jw}$ to represent the particle flux at the wall, and $\Phi_{jw}$ to represent the particle current flux at the wall, then the current flux is zero $\sum_j \Phi_{jw} = 0$ because the total charge is conserved at the wall, namely,

$$\sigma_{pw} = \sigma_{sw} + \sigma_{iw}, \tag{7}$$

where the particle fluxes are $\sigma_j = n_{j0}v_{j0} = n_j v_j$ .

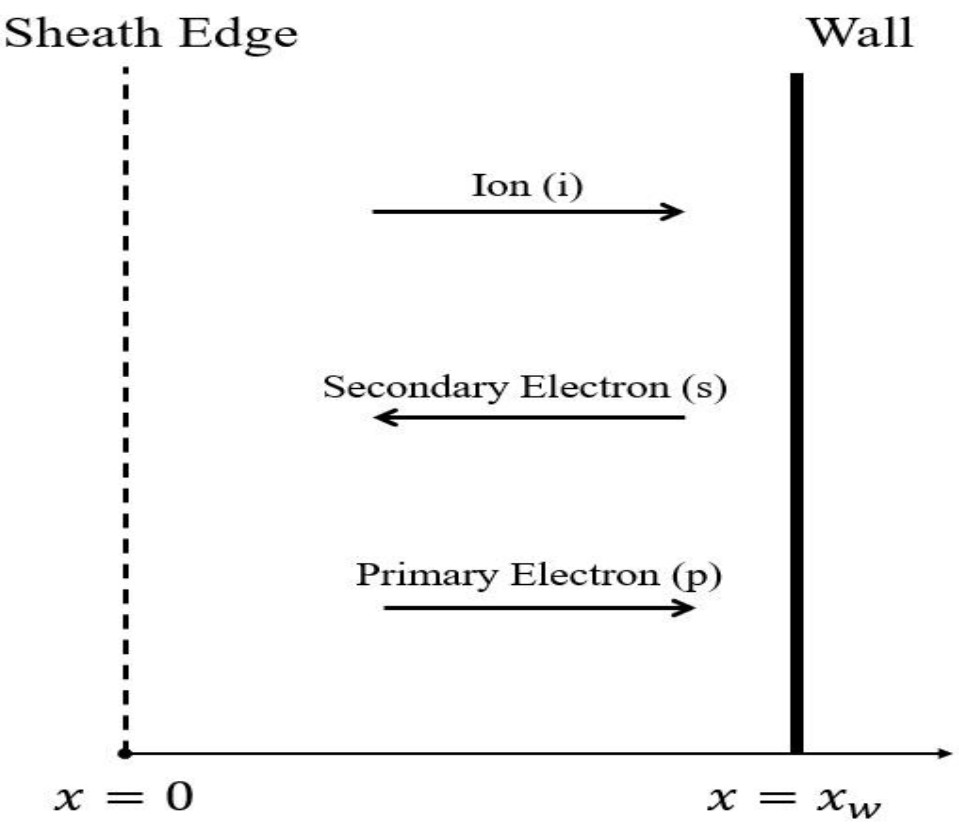

Figure 1. Schematic diagram of plasma sheath

If the secondary electron emission coefficient $\gamma$ at the wall is defined as $\gamma=\sigma_{sw}/\sigma_{pw}$, combining Eq. (7) we get

$$\sigma_{pw}=\frac{1}{1-\gamma}\sigma_{iw}, \qquad \sigma_{sw}=\frac{\gamma}{1-\gamma}\sigma_{iw}. \tag{8}$$

From Eqs.(2) and (3) we can get

$$v_s=\sqrt{{v_{sw}}^2+\frac{2}{m_e}\left(\varphi_w-\varphi\right)}, \tag{9}$$

where $\varphi_w$ is the floating potential at the wall. From Eqs. (6) - (9), the particle number density of the secondary electron is obtained as

$$n_s=\frac{n_{sw}v_{sw}}{v_s}=\frac{\gamma\sigma_{iw}}{(1-\gamma)v_s}=\frac{\gamma n_{io}v_{io}}{(1-\gamma)v_s}=\frac{\gamma n_{io}}{1-\gamma}\sqrt{\frac{m_e v_{i0}^2}{2e\left(\varphi-\varphi_w\right)+m_e v_{sw}^2}}. \tag{10}$$

Conventionally, if the primary electrons is assumed to be a Maxwellian distribution, their number density is $n_p = n_{p0}\exp(e\varphi/kT_p)$. And then, by using the quasi-neutral condition at the boundary of the sheath, $n_{p0}+n_{s0}=n_{i0}$, and Eqs.(6) and (10), one can derive that

$$n_p=\left(n_{i0}-n_{s0}\right)\exp\left(\frac{e\varphi}{kT_p}\right)=n_{i0}\left(1-\frac{\gamma}{1-\gamma}\sqrt{\frac{m_e v_{i0}^2}{m_e v_{sw}^2-2e\varphi_w}}\right)\exp\left(\frac{e\varphi}{kT_p}\right). \tag{11}$$

For the Poisson equation,

$$\frac{d^2\varphi}{dx^2}=\frac{e}{\varepsilon_0}\left(n_p+n_s-n_i\right), \tag{12}$$

if one makes a dimensionless transformation by using the following six parameters,

$$\xi=\frac{x}{\lambda_D},\ \theta_j=\frac{v_j}{u_i}, N_j=\frac{n_j}{n_{i0}},\ \psi=\frac{e\varphi}{kT_p},\ \mu=\frac{m_i}{m_e},\ M^2=\frac{m_i v_{i0}^2}{kT_p}, \tag{13}$$

where $\lambda_D=\sqrt{\varepsilon_0 kT_p/n_{i0}e^2}$ is the Debye length, $M$ is the Mach number, $u_i=\sqrt{kT_p/m_i}$ is the ion acoustic speed, and $T_p$ is the primary electron temperature, then one has that

$$\frac{d^2\psi}{d\xi^2}=N_p+N_s-N_i, \tag{14}$$

where the dimensionless particle population densities are, respectively,

$$N_i=\sqrt{1\Big/\left(1-\frac{2\psi}{M^2}\right)}, \tag{15}$$

$$N_s=\frac{\gamma}{1-\gamma}\sqrt{\frac{M^2}{2\mu\left(\psi-\psi_w\right)+\theta_{sw}^2}}, \tag{16}$$

$$N_p=\left(1-\frac{\gamma}{1-\gamma}\sqrt{\frac{M^2}{\theta_{sw}^2-2\mu\psi_w}}\right)\exp\left(\psi\right), \tag{17}$$

where $\theta_{sw} = v_{sw}/u_i$ the dimensionless initial speed of SEE and $\psi_w$ is the floating potential at the wall.

Usually, one uses the Sagdeev potential to analyze the properties of quasi-particles. The Sagdeev potential $V$ is defined [4] by

$$\frac{d^2\psi}{d\xi^2} = -\frac{dV(\psi)}{d\psi}. \tag{18}$$

The configuration of Sagdeev potential must be a potential well, therefore the function value and first derivative of Sagdeev potential at $\psi$=0 are both zero, and the second derivative of Sagdeev potential must be less than zero at $\psi$=0. One has that

$$\left(\frac{d^2V}{d\psi^2}\right)_{\psi=0} \le 0. \tag{19}$$

And then, on the basis of Eqs.(14)-(19), one obtain the Bohm criterion[17],

$$M^2 \ge \frac{1}{1-\frac{\gamma}{1-\gamma}\sqrt{\frac{M^2}{\theta_{sw}^2 - 2\psi_w\mu}} - \mu\frac{\gamma}{1-\gamma}\sqrt{\frac{M^2}{\left(\theta_{sw}^2 - 2\psi_w\mu\right)^3}}}. \tag{20}$$

Eq. (20) is an implicit inequality with the Mach number $M$ appearing on both sides. When the equality holds, it defines the critical Mach number $M_c$ that satisfies the sheath formation condition, and its value has to be determined numerically. In the limit of no secondary electron emission ($\gamma \to 0$), this inequality reduces to the classical Bohm criterion $M \ge 1$.

In the above basic theory, the ions and the electrons (primary electrons and secondary electrons) in the plasma sheath were often assumed to be a Maxwellian velocity distribution. But in many plasma situations such as astrophysical and space plasmas, non-Maxwellian velocity distributions of the particles are very common, among which there is a non-thermal velocity $\alpha$-distribution. In 1995, Cairns et al introduced a velocity $\alpha$-distribution[19] inspired by the observation data of Freja and Viking satellites, called as Cairns-distribution, which is non-thermal, written by

$$f_\alpha(v) = C_\alpha\left(1+\alpha\frac{m^2v^4}{k^2T^2}\right)\exp\left(-\frac{mv^2}{2kT}\right), \tag{21}$$

where $T$ is temperature, $k$ is Boltzmann constant, $m$ is mass of the particle, $\alpha \neq 0$ is a non-thermal parameter which describes the amount of high-energy particles in the plasma, and $C_\alpha$ is a normalization constant,

$$C_\alpha = \frac{n}{3\alpha+1}\sqrt{\frac{m}{2\pi kT}}. \tag{22}$$

If one takes $\alpha = 0$, the distribution function (21) returns to a Maxwellian distribution.

## III. The plasma sheath with the Cairns-distribution

### *3.1. The generalized Bohm criterion*

In the plasma sheath model, we assume the primary electrons to be the Cairns-distribution. And when there is the potential $\varphi(x)$ the Cairns-distribution (21) can be written by[19]

$$f_{p,\alpha}(v_p, x) = C_\alpha\left[1+\frac{\alpha\left(m_ev_p^2 - 2e\varphi\right)^2}{k^2T_p^2}\right]\exp\left(\frac{m_ev_p^2 - 2e\varphi}{2kT_p}\right), \tag{23}$$

where the normalization constant is

$$C_\alpha = \frac{n_{p0}}{3\alpha+1}\sqrt{\frac{m_e}{2\pi kT_p}} . \tag{24}$$

Based on this distribution function, number density of the primary electrons is derived as

$$n_{p,\alpha}\left(x\right) = n_{p0}\left[1 - A_\alpha\frac{e\varphi}{kT_p} + A_\alpha\left(\frac{e\varphi}{kT_p}\right)^2\right]\exp\left(\frac{e\varphi}{kT_p}\right), \tag{25}$$

where $A_\alpha$=4$\alpha$/(3$\alpha$+1). Correspondingly, the dimensionless particle density (17) becomes

$$N_{p,\alpha} = \left(1 - \frac{\gamma}{1-\gamma}\sqrt{\frac{M_\alpha^2}{\theta_{sw}^2 - 2\mu\psi_w}}\right)\left(1 - A_\alpha\psi + A_\alpha\psi^2\right)\exp\left(\psi\right). \tag{26}$$

On the basis of Eqs.(18)-(19) as well as Eqs.(13)-(16) and Eq.(26), one can obtain the following inequation,

$$-\frac{d}{d\psi}\left[\left(1 - \frac{\gamma}{1-\gamma}\sqrt{\frac{M_\alpha^2}{\theta_{sw}^2 - 2\mu\psi_w}}\right)\left(1 - A_\alpha\psi + A_\alpha\psi^2\right)\exp(\psi) + \frac{\gamma}{1-\gamma}\sqrt{\frac{M_\alpha^2}{2\mu(\psi-\psi_w)+\theta_{sw}^2}} - \sqrt{\frac{M_\alpha^2}{M^2 - 2\psi}}\right]_{\psi=0} \le 0 \cdot \tag{27}$$

And then, from (27) one can derive a generalized Bohm criterion for the plasma sheath if the primary electrons has the Cairns-distribution, that is

$$M_\alpha^2 \ge \frac{1}{\left(1 - A_\alpha\right)\left(1 - \frac{\gamma}{1-\gamma}\sqrt{\frac{M_\alpha^2}{\theta_{sw}^2 - 2\psi_w\mu}}\right) - \frac{\mu\gamma}{1-\gamma}\sqrt{\frac{M_\alpha^2}{\left(\theta_{sw}^2 - 2\psi_w\mu\right)^3}}} . \tag{28}$$

Obviously, this Bohm criterion depends on the non-thermal $\alpha$-parameter, when one takes $\alpha = 0$ , one has $A_\alpha$=0 and thus the generalized Bohm criterion (28) returns to the classical form (20). From the new criterion (28), we obtain a new equation for the critical Mach number $M_{c,\alpha}$ (i.e., the Bohm speed) with respect to the SEE coefficient $\gamma$ and the floating potential $\psi_w$ at the wall,

$$M_{c,\alpha}^2\left[\left(1 - A_\alpha\right)\left(1 - \frac{\gamma}{1-\gamma}\sqrt{\frac{M_{c,\alpha}^2}{\theta_{sw}^2 - 2\psi_w\mu}}\right) - \frac{\mu\gamma}{1-\gamma}\sqrt{\frac{M_{c,\alpha}^2}{\left(\theta_{sw}^2 - 2\psi_w\mu\right)^3}}\right] = 1 . \tag{29}$$

This critical Mach number also depends on the non-thermal $\alpha$-parameter, and we find that the generalized Bohm criterion can be written as $M_\alpha \ge M_{c,\alpha}$ .

*3.2. The floating potential at the wall*

The concept of floating potential was once used to describe the potential on Langmuir probes. Because the probe is suspended into the plasma for detection, the potential energy on the suspended probe is also called suspended potential. When the net current flux on the probe surface is zero, a balance is achieved between the incoming current and the outgoing current, and then the potential energy state on the probe is in a suspended state. In fact, the potential on various objects such as electrodes, walls, workpieces and dust grains that are immersed in or come into contact with the plasma with zero net electricity may be called the floating potential[20, 21].

In order to study the generalized Bohm criterion (28), we need to analyse the floating potential $\psi_w$ at the wall. At the same time, as the important part of the sheath potential structure, the property of floating potential at the wall is also very important. In the equilibrium condition (7) between the secondary electron flux $\sigma_{sw}$, the primary electron flux $\sigma_{pw}$ and the ion flux $\sigma_{iw}$ at the wall, the primary electron flux $\sigma_{pw}$ at the wall is calculated on the basis of the Cairns $\alpha$-distribution (23) by

$$\sigma_{pw}=\int_0^{\infty} f_{p,\alpha}\left(v_p,x\right)dv_p=C_\alpha\int_0^{\infty}dv_p\left[1+\frac{\alpha\left(m_e v_p^2-2e\varphi_w\right)^2}{k^2T_p^2}\right]\exp\left(-\frac{m_e v_p^2-2e\varphi_w}{2kT_p}\right)v_p$$

$$=\frac{n_{p0}}{\left(3\alpha+1\right)}\sqrt{\frac{kT_p}{2\pi m_e}}\exp\left(\frac{e\varphi_w}{kT_p}\right)\left(1+8\alpha-8\alpha\frac{e\varphi_w}{kT_p}+4\alpha\frac{e^2\varphi_w^2}{k^2T_p^2}\right). \tag{30}$$

The ions flux and the secondary electron flux are respectively,

$$\sigma_{iw}=n_{iw}v_{iw}=n_{i0}v_{i0}\ \text{ and }\ \sigma_{sw}=\frac{\gamma}{1-\gamma}n_{i0}v_{i0}. \tag{31}$$

Then, based on the charge conservation Eq. (7) we get that

$$\frac{n_{p0}}{\left(3\alpha+1\right)}\sqrt{\frac{kT_p}{2\pi m_e}}\left(1+8\alpha-8\alpha\frac{e\varphi_w}{kT_p}+4\alpha\frac{e^2\varphi_w^2}{k^2T_p^2}\right)\exp\left(\frac{e\varphi_w}{kT_p}\right)=\frac{n_{i0}v_{i0}}{1-\gamma}. \tag{32}$$

By using the dimensionless transformation (13), Eq.(32) is written as

$$\sqrt{\frac{\mu}{2\pi}}\left(1-\frac{\gamma}{1-\gamma}\frac{M_\alpha}{\sqrt{\theta_{sw}^2-2\mu\psi_w}}\right)\left(\frac{1}{3\alpha+1}+2A_\alpha-2A_\alpha\psi_w+A_\alpha\psi_w^2\right)\exp\left(\psi_w\right)=\frac{M_\alpha}{1-\gamma}. \tag{33}$$

This equation gives a relation between the floating potential at the wall and the other plasma physical quantities of the sheath if the primary electrons are non-thermal and have the Cairns $\alpha$-distribution. This new relation depends on the non-thermal $\alpha$-parameter, only when one takes $\alpha=0$, it returns the original relation when the primary electrons were assumed to be a Maxellian velocity distribution [17],

$$\sqrt{\frac{\mu}{2\pi}}\left(1-\frac{\gamma}{1-\gamma}\frac{M}{\sqrt{\theta_{sw}^2-2\mu\psi_w}}\right)\exp\left(\psi_w\right)=\frac{M}{1-\gamma}. \tag{34}$$

### *3.3. The critical SEE coefficient*

We now study the critical SEE coefficient $\gamma_c$ in the plasma sheath with the Cairns-distributed primary electrons. Hobbs and Wesson obtained an expression of the critical SEE coefficient [6], $\gamma_c\approx 1-8.3\sqrt{m_e/m_i}$ . If SEE effect is weak, then $\gamma<\gamma_c$ ; if SEE effect is strong, then $\gamma>\gamma_c$ ; and if SEE effect is moderate, then $\gamma=\gamma_c$ .As we know that [5] when the SEE effect is weak, the SEE coefficient $\gamma<\gamma_c$ , at this time, the potential distribution type of the sheath is a classical sheath and the floating potential at the wall has a negative potential relative to the plasma. The electric field near the wall points to the wall, so all the secondary electrons emitted from the wall will enter the sheath and do not return. When the SEE effect is strong, the SEE coefficient $\gamma>\gamma_c$, at this time, the potential distribution type of the sheath is an inverse sheath. The floating potential of the wall has a positive potential relative to the plasma. And the electric field near the wall points to the plasma region, and most of the secondary electrons will return to the wall. When the SEE effect is moderate, the SEE coefficient $\gamma=\gamma_c$, at this time, the potential distribution type of sheath is a space charge limited (SCL) transition sheath, and the sheath structure is that its electric field near the wall is zero. $\gamma_c$ is called the critical SEE coefficient. Mathematically, the SCL transition sheath corresponds to the special form of wall electric field being zero, and physically, the corresponding SEE effect intensity is relatively moderate. Therefore, we take the SCL transition sheath structure to study the critical SEE coefficient.

In order to find the critical SEE coefficient $\gamma_c$ , by making use of the normalized form of the Poisson equation (14), one finds that

$$\frac{d\psi}{d\xi} d\left(\frac{d\psi}{d\xi}\right) = (N_{p,\alpha} + N_s - N_i) d\psi \, . \tag{35}$$

Making integral from $\psi_0$ to $\psi_w$ on both sides of Eq.(35), one has

$$\int_{E_0}^{E_w} \frac{d\psi}{d\xi} d\left(\frac{d\psi}{d\xi}\right) = \int_{\psi_0}^{\psi_w} (N_p + N_s - N_i) d\psi \, , \tag{36}$$

where the electric field at the boundary of sheath is $E_0$=0 and $\psi_0$=0. In the SCL transition sheath, the electric field $E_w$=0 at the wall (i.e., $d\psi/d\xi = 0$ at the wall), $\gamma = \gamma_c$ , and so Eq.(36) becomes

$$\int_0^{\psi_w} (N_{p,\alpha} + N_s - N_i) d\psi = 0 \, . \tag{37}$$

Substituting Eqs.(15), (16) and (27) in to Eq.(37), one derives the equation of the critical SEE coefficient $\gamma_c$ when the plasma sheath has the Cairns-distributed primary electrons,

$$\begin{aligned} &\left(1 - \frac{M_\alpha \gamma_c}{(1-\gamma_c)\sqrt{\theta_{sw}^2 - 2\psi_w \mu}}\right) \left[ A_\alpha \left( \psi_w^{\,2} - 3\psi_w + 3 \right) \exp\left(\psi_w\right) - 3A_\alpha + \exp\left(\psi_w\right) - 1 \right] \\ &+ \frac{M_\alpha \gamma_c}{(1-\gamma_c)\mu} \left( \theta_{sw} - \sqrt{\theta_{sw}^2 - 2\psi_w \mu} \right) + M_\alpha \left( \sqrt{M^2 - 2\psi_w} - M_\alpha \right) = 0. \end{aligned} \tag{38}$$

This new equation depends on the non-thermal $\alpha$-parameter, which can be used to solve the new critical SEE coefficient $\gamma_c$. When we take $\alpha = 0$ , it returns to the form if the primary electrons were a Maxwellian distribution[22], namely,

$$\begin{aligned} &\left(1 - \frac{M\gamma_c}{(1-\gamma_c)\sqrt{\theta_{sw}^2 - 2\psi_w \mu}}\right) \left(\exp\left(\psi_w\right) - 1\right) + \frac{M\gamma_c}{\left(1-\gamma_c\right)\mu} \left(\theta_{sw} - \sqrt{\theta_{sw}^2 - 2\psi_w \mu}\right) \\ &+ M\left(\sqrt{M^2 - 2\psi_w} - M\right) = 0. \end{aligned} \tag{39}$$

## IV. Numerical analyses

In order to show the properties of the plasma sheath being the non-thermal primary electrons with the Cairns $\alpha$-distribution more clearly, we now make numerical analyses on the basis of the generalized Bohm criterion (28), the new floating potential (33) and the new critical SEE coefficient (38) of the plasma sheath. In numerical analyses, as an example, the plasma characteristic data are taken in an argon plasma containing secondary emission electrons [23], such as the mass ratio of an ion to an electron $\mu$= 7.3334×10$^4$, the dimensionless exit speed of secondary electrons $\theta_{sw}$ = 20, the Mach number $M_\alpha$=1.4, the floating potential at the wall $\psi_w$ = –3, and the SEE coefficient $\gamma$ = 0.4. In addition, according to the Bohm criterion, $M_\alpha$ should be taken greater than the critical value $M_{c,\alpha}$ (the Bohm speed) when it is used as an independent variable.

### *4.1. The generalized Bohm criterion and Bohm speed*

The numerical analyses are performed for the critical Bohm speed $M_{c,\alpha}$ based on Eq. (29). In Fig. 2a and Fig. 2b, we illustrate the Bohm speed $M_{c,\alpha}$ as a function of the non-thermal $\alpha$-parameter for three representative values of the SEE coefficient and the floating potential, respectively. Figure 2a shows results for three different SEE coefficients $\gamma$ = 0.1, 0.5 and 0.9, while Fig. 2b presents results for three different floating potentials $\psi_w$ = -4, -3 and -2.

It is shown that the Bohm speed depends significantly on the non-thermal $\alpha$-parameter, but it is almost unaffected by the SEE coefficient $\gamma$ and the floating potential at the wall $\psi_w$, as evidenced by the near overlap of the curves in both panels. The Bohm speed $M_{c,\alpha}$ increases substantially with increasing $\alpha$-parameter, where $\alpha$ = 0 corresponds to the classical case of a Maxwellian distribution. Thus, the Bohm speed $M_{c,\alpha}$ in a plasma sheath with non-thermal primary electrons described by the Cairns distribution is generally larger than that with a Maxwellian distribution.

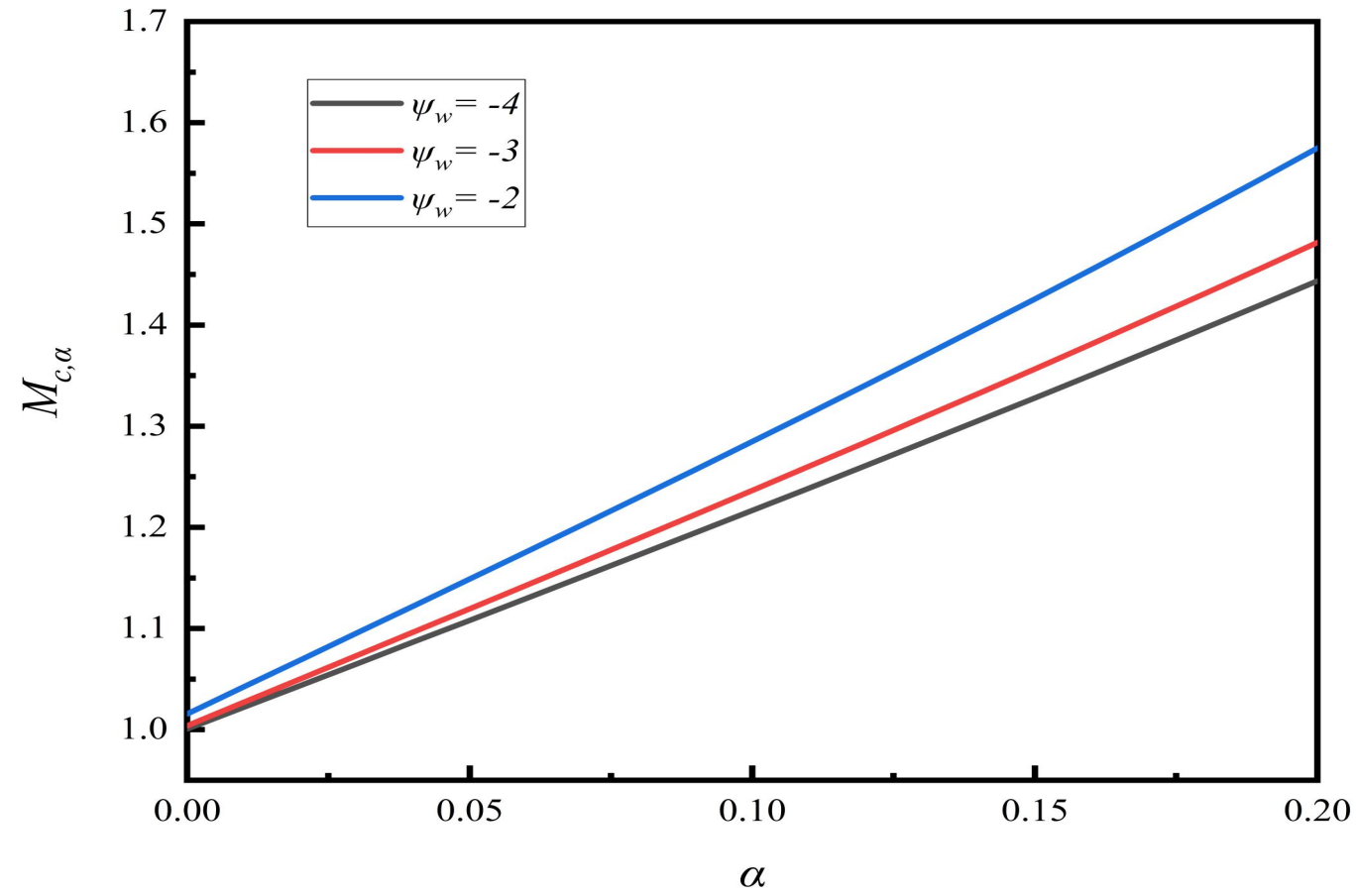

**Figure 2a.** The generalized Bohm speed as a function of the Cairns $\alpha$-parameter for three different $\psi_w$-parameters.

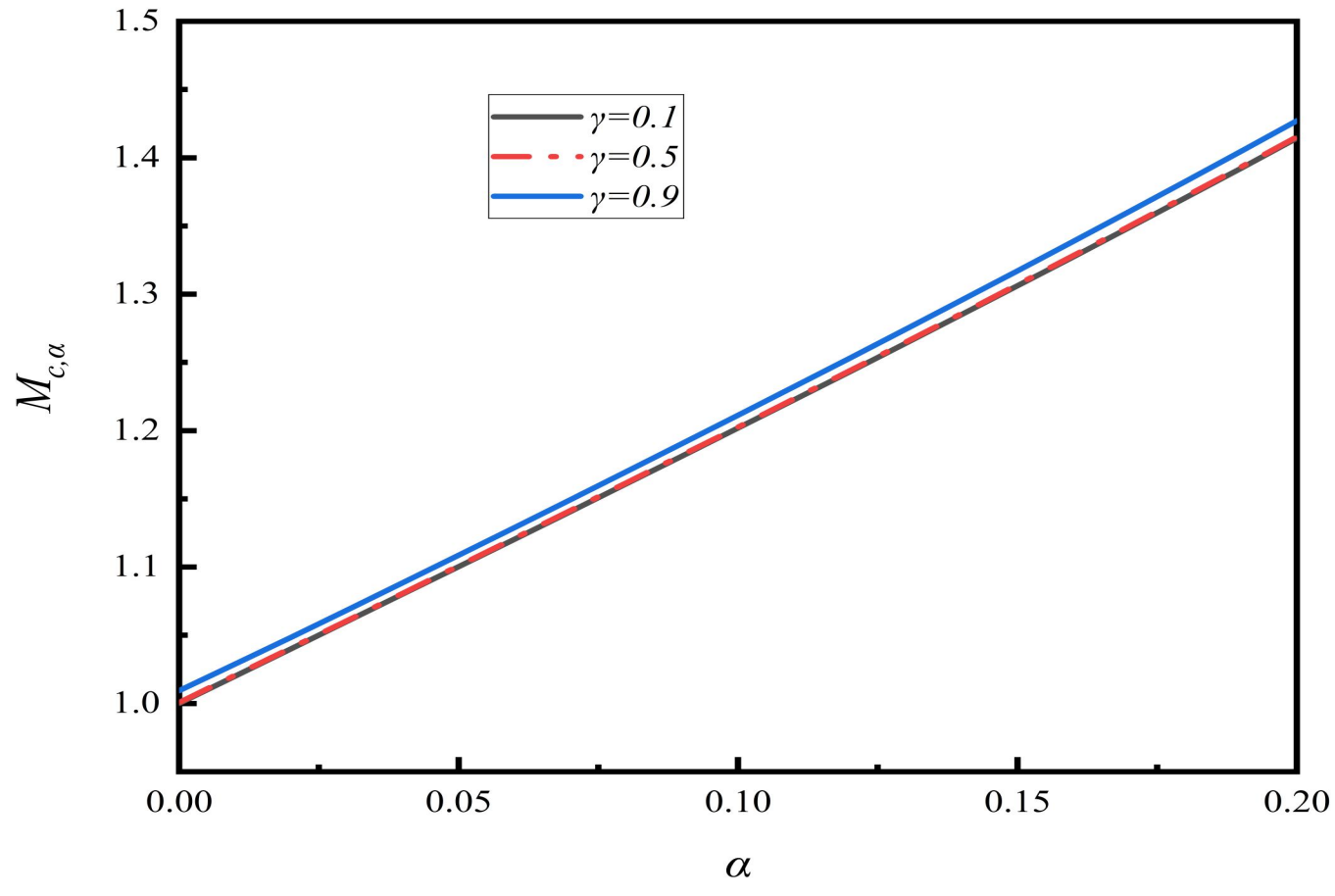

**Figure 2b.** The generalized Bohm speed as a function of the Cairns $\alpha$-parameter for three different $\gamma$-parameters.

Physically, an increase in the non-thermal $\alpha$-parameter leads to a higher population of electrons in energetic states and the appearance of non-monotonic features in the velocity distribution. This enhances the electron flux toward the wall and allows more electrons to accumulate in the sheath region. Consequently, ions require greater kinetic energy to satisfy the generalized Bohm criterion at the sheath entrance, resulting in an increased Bohm speed.

*4.2. The new floating potential at the wall*

The numerical analyses are made of the floating potential at the wall $\psi_w$ based on Eq.(33). In Fig.3a and Fig.3b, we illustrated the floating potential $\psi_w$ as a function of the SEE coefficient $\gamma$ and the Mach number $M_\alpha$, respectively, for three different values of the $\alpha$-parameter, $\alpha = 0$, 0.05, 0.15. In Fig.3b, the starting value of Mach number $M_\alpha$ is taken as the critical value $M_{c,\alpha}$ to ensure the Bohm criterion $M_\alpha \geq M_{c,\alpha}$. It is shown that the floating potential depend significantly on the non-thermal $\alpha$-parameter, and it decreases with increase of the non-thermal $\alpha$-parameter, where $\alpha = 0$ is the case of the plasma with a Maxwellian distribution. Thus, the floating potential in the plasma sheath having the non-thermal primary electrons with the Cairns-distribution is generally less than that with a Maxwellian distribution.

In Fig.3a, we show that the floating potential $\psi_w$ increase slowly with increase of the SEE coefficient $\gamma$, but in Fig.3b, we show that it is almost unchanged with increase of the Mach number $M_\alpha$.

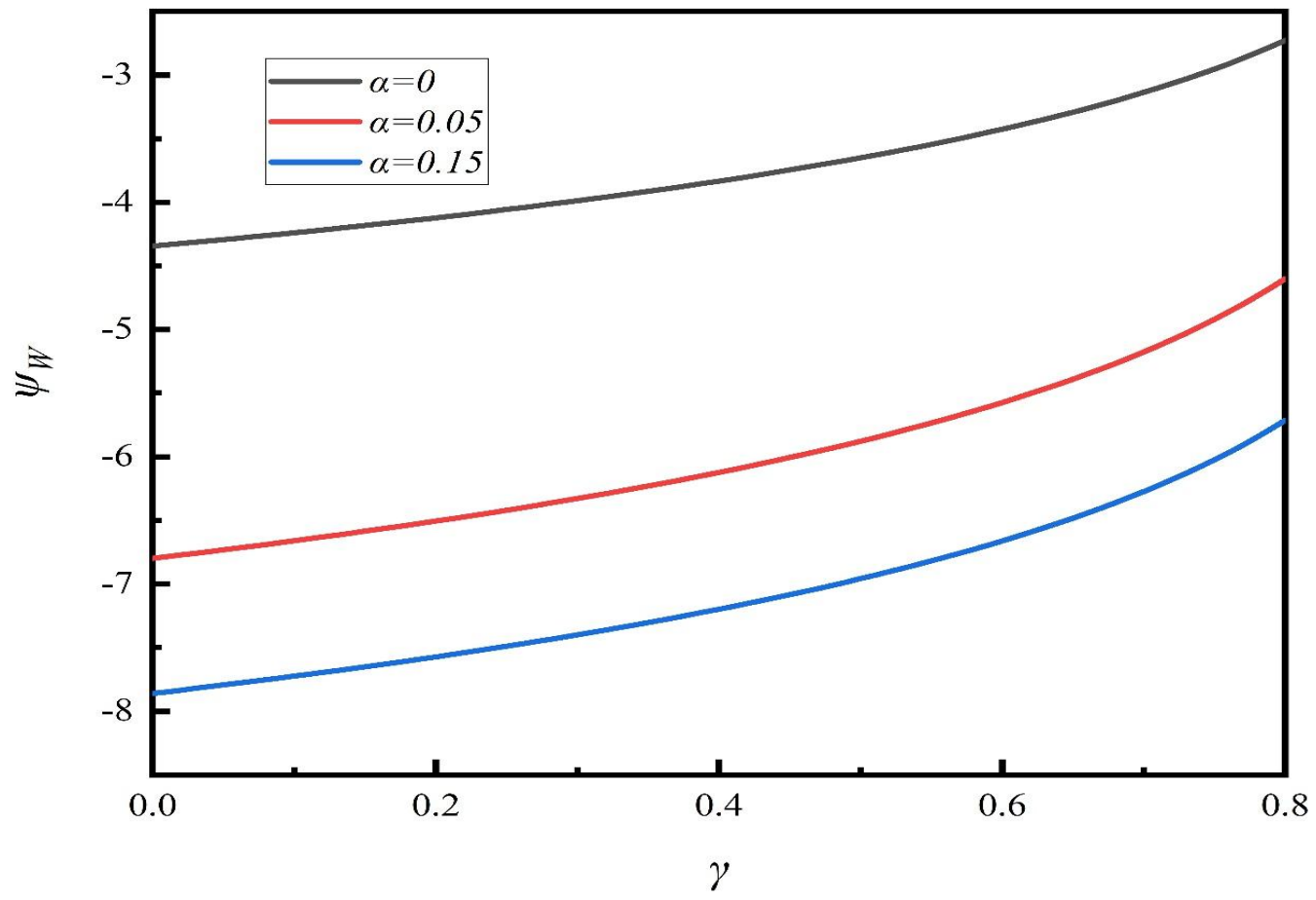

**Figure 3a.** The new floating potential as a function of the SEE coefficient for three different $\alpha$-parameters.

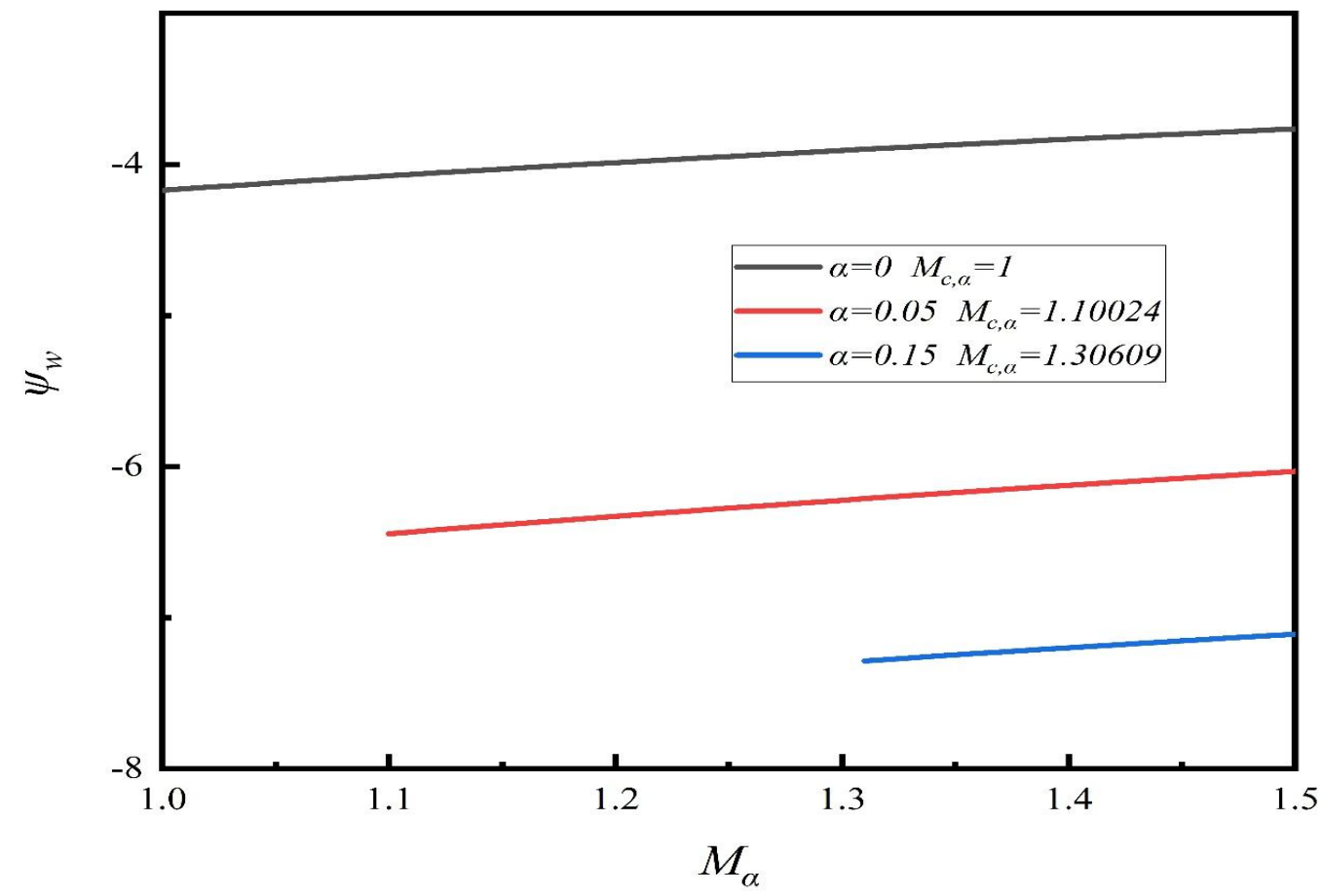

**Figure 3b.** The new floating potential as a function of the Mach number for three different $\alpha$-parameters.

In the Figs. we observed that the floating potential decreases (i.e., it becomes more negative) with increase of the non-thermal parameter $\alpha$. This may be understood through the following mechanism. With increase of the non-thermal parameter $\alpha$, the primary electrons in the high-energy tails are enhanced. This increases significantly the flux of the energetic electrons that can reach the wall against the retarding sheath potential. To maintain the fundamental condition of zero net current at the wall, the system has to compensate for this increased primary electron current by adjusting the floating potential to a more negative value. This adjustment restores balance in two key ways: (1) it attracts a greater flux of positive ions, and (2) it reflects the lower-energy portion of the primary electron population more effectively. The net effect reaches a new equilibrium where the increased influx of energetic electrons is counteracted, establishing a steady-state floating potential with a lower (more negative) value.

*4.3. The new critical SEE coefficient*

The numerical analyses are made of the critical SEE coefficient $\gamma_c$ based on Eq.(38). In Fig.4a and 4b, we illustrated the critical SEE coefficient $\gamma_c$ as a function of the floating potential at the wall $\psi_w$ and the Mach number $M_\alpha$, respectively, for three different values of the non-thermal $\alpha$-parameter, $\alpha = 0, 0.05, 0.15$. It is shown that the critical SEE coefficient depends significantly on the $\alpha$-parameter, and it decreases with the increase of the $\alpha$-parameter, where $\alpha = 0$ is the case of the plasma with a Maxwellian distribution. Thus, the critical SEE coefficient in the plasma sheath having the non-thermal primary electrons with the Cairns-distribution is generally more than that with a Maxwellian distribution.

In Fig.4a, we show that the critical SEE coefficient $\gamma_c$ decreases with increase of the floating potential $\psi_w$, but the speed at which it decreases depends significantly on the $\alpha$-parameter. When $\alpha = 0$, $\gamma_c$ decreases very slowly with increase of $\psi_w$, but with increase of the $\alpha$-parameter, $\gamma_c$ become to decrease quickly with increase of $\psi_w$

In Fig.4b, we show that the critical SEE coefficient $\gamma_c$ increases with increase of the Mach number $M_\alpha$, but the speed at which it increases depends significantly on the $\alpha$-parameter. When $\alpha = 0$, $\gamma_c$ increases very slowly with increase of $M_\alpha$, but with increase of the $\alpha$-parameter, $\gamma_c$ become to increase quickly with increase of $M_\alpha$.

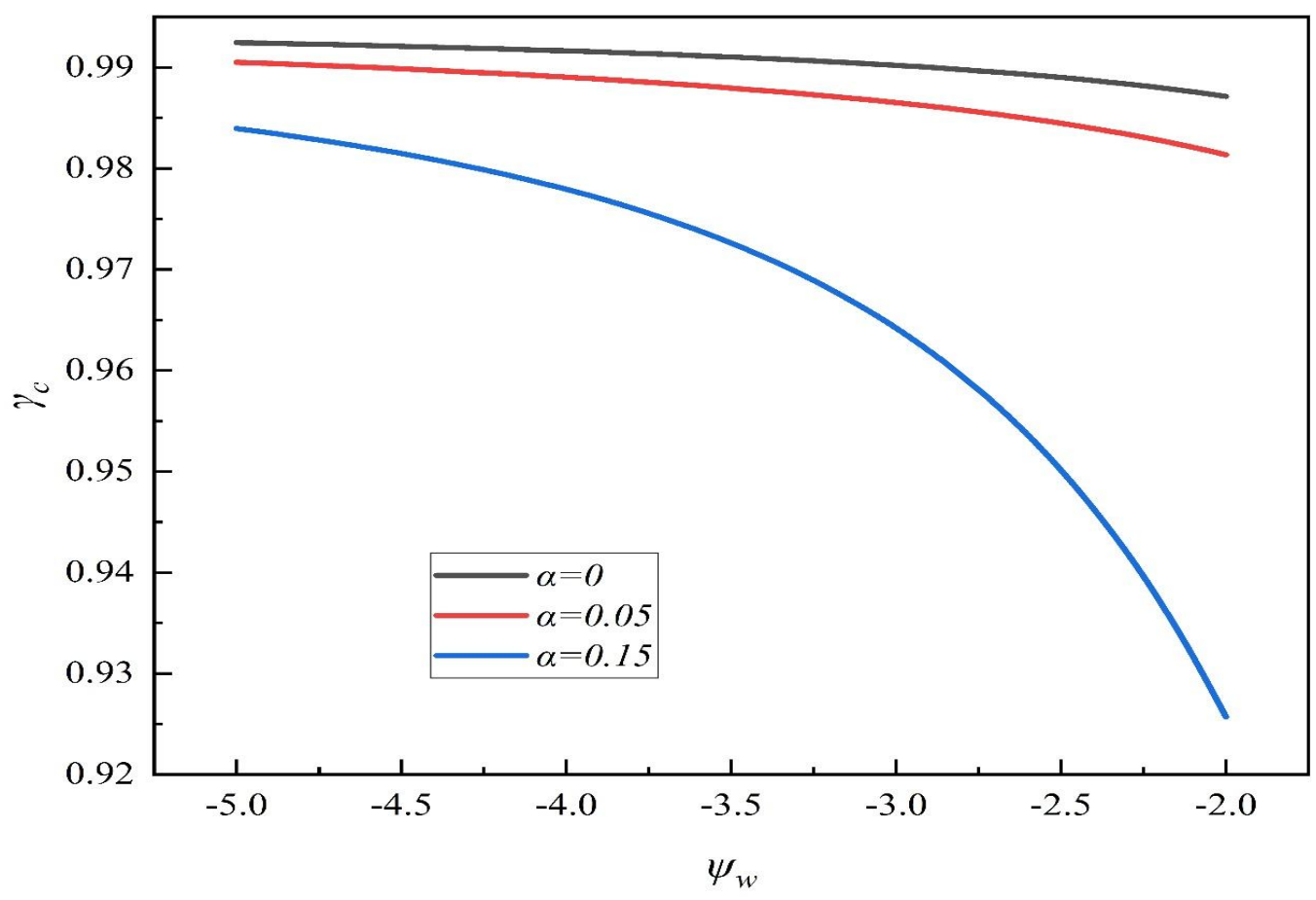

**Figure 4a.** The new critical SEE coefficient as a function of the floating potential for three different values of the $\alpha$-parameter.

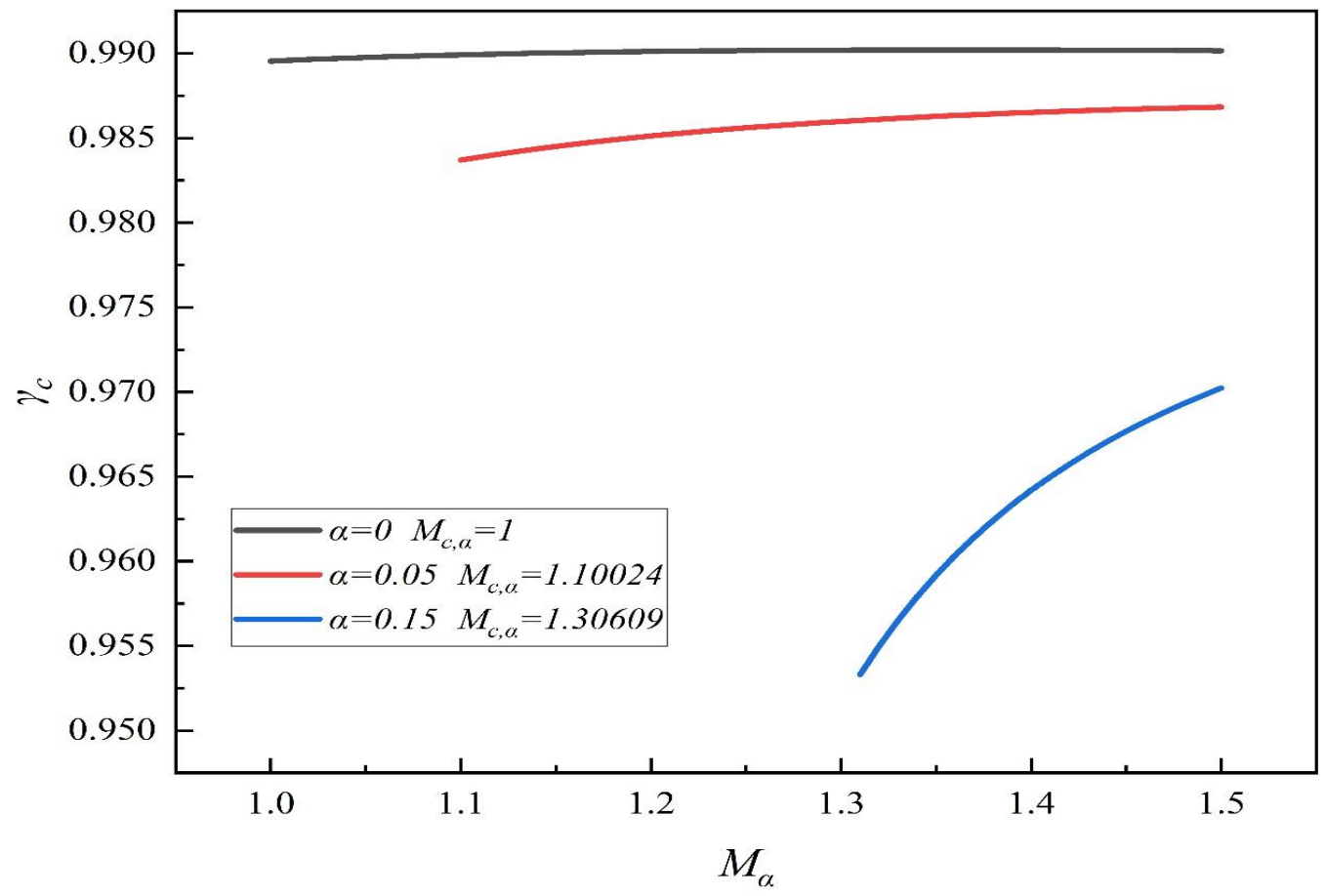

**Figure 4b**. The new critical SEE coefficient as a function of the Mach number for three different values of the $\alpha$-parameters.

## V. Conclusions

In conclusion, we have studied the properties of the plasma sheath with secondary electrons and non-thermal primary electrons with a Cairns-distribution, including the Bohm criterion, the floating potential and the SEE coefficient. We have derived the expression of a generalized Bohm criterion given by the inequality (28), the equation (33) of the floating potential $\psi_w$ at the wall and the equation (38) of the critical SEE coefficient $\gamma_c$ when the plasma sheath having the non-thermal primary electrons with Cairns $\alpha$-distribution. We also obtained the equation (29) of the Bohm speed $M_{c,\alpha}$. These new relations are all related to the non-thermal $\alpha$-parameter. When we take $\alpha = 0$, they all return to those forms in the plasma sheath with the Maxwell-distributed primary electrons.

We have made the numerical analyses to illustrate the Bohm speed, the floating potential at the wall and the critical SEE coefficient, respectively, for three different values of the $\alpha$-parameter. The results showed that

(a) The Bohm speed depend $M_{c,\alpha}$ significantly on the non-thermal $\alpha$-parameter, but it is almost unaffected by the SEE coefficient and the floating potential at the wall. The Bohm speed increases with increase of the $\alpha$-parameter, so the Bohm speed of the plasma sheath having the non-thermal primary electrons with the Cairns-distribution is generally more than that with a Maxwellian distribution.

(b) The floating potential $\psi_w$ at the wall depend significantly on the non-thermal $\alpha$-parameter, and it decreases with increase of the non-thermal $\alpha$-parameter. Therefore, the floating potential of the plasma sheath having the non-thermal primary electrons with the Cairns-distribution is generally less than that with a Maxwellian distribution. The floating potential increase slowly with increase of the SEE coefficient $\gamma$, but it is almost unchanged with increase of the Mach number $M_\alpha$.

(c) The critical SEE coefficient $\gamma_c$ depends significantly on the $\alpha$-parameter, and it decreases with the increase of the $\alpha$-parameter. The critical SEE coefficient in the plasma sheath having the non-thermal primary electrons with the Cairns-distribution is generally more than that with a Maxwellian distribution. The critical SEE coefficient decreases with increase of the floating potential $\psi_w$, but the speed at which it decreases depends significantly on the $\alpha$-parameter. When $\alpha = 0$, $\gamma_c$ decreases very slowly with increase of $\psi_w$, but with increase of the $\alpha$-parameter, $\gamma_c$ become to decrease quickly with increase of $\psi_w$. The critical SEE coefficient increases with increase of the Mach number $M_\alpha$, but the speed at which it increases depends significantly on the $\alpha$-parameter. When $\alpha = 0$, $\gamma_c$ increases very slowly with increase of $M_\alpha$, but with increase of the $\alpha$-parameter, $\gamma_c$ become to increase quickly with increase of $M_\alpha$ .